\documentclass[twocolumn]{aastex63}

\usepackage{xspace}
\usepackage{amsmath}
\usepackage{hyperref}
\usepackage[figuresright]{rotating}
\usepackage{enumerate}
\usepackage{bm}

\defcitealias{fm07}{FM07}
\defcitealias{g12}{G12}

\newcommand{\ctfm}{\citetalias{fm07}\xspace}

\newcommand{\ctgud}{\citetalias{g12}\xspace}

\newcommand{\hi}{\ion{H}{1}\xspace}
\newcommand{\htwo}{H$_2$\xspace}
\newcommand{\nhi}{$N(\text{\hi})$\xspace}
\newcommand{\nhtwo}{$N(\text{\htwo})$\xspace}
\newcommand{\nh}{$N(\text{H})$\xspace}
\newcommand{\nhplus}{$N(\text{\hi} + \text{\htwo})$\xspace}
\newcommand{\alam}{$A_\lambda$\xspace}

\newcommand{\ebv}{$E(B-V)$\xspace}
\newcommand{\chisq}{$\chi^2$\xspace}
\newcommand{\redchisq}{$\chi^2_\text{red}$\xspace}
\newcommand{\bnir}{$\beta_\text{NIR}$\xspace}
\newcommand{\feii}{\ion{Fe}{2}\xspace}
\newcommand{\oi}{\ion{O}{1}\xspace}
\newcommand{\ovi}{\ion{O}{6}\xspace}
\newcommand{\mgii}{\ion{Mg}{2}\xspace}
\newcommand{\caii}{\ion{Ca}{2}\xspace}
\newcommand{\nfeii}{$N(\text{\feii})$\xspace}
\newcommand{\noi}{$N(\text{\oi})$\xspace}
\newcommand{\novi}{$N(\text{\ovi})$\xspace}
\newcommand{\nmgii}{$N(\text{\mgii})$\xspace}
\newcommand{\ncaii}{$N(\text{\caii})$\xspace}

\newcommand{\um}{$\mu$m\xspace}
\newcommand{\invum}{$\mu$m$^{-1}$\xspace}
\newcommand{\Rv}{$R_V$\xspace}
\newcommand{\Av}{$A_V$\xspace}
\newcommand{\avnh}{$A_V/N(\text{H})$\xspace}
\newcommand{\nhav}{$N(\text{H})/A_V$\xspace}
\newcommand{\nhebv}{$N(\text{H})/E(B-V)$\xspace}
\newcommand{\alamnh}{$A_\lambda/N(\text{H})$\xspace}

\newcommand{\Abest}{$A_{2900}$\xspace}

\newcommand{\nc}{$N(\text{C})$\xspace}
\newcommand{\nd}{$N_\text{d}$\xspace}

\newcommand{\lognhplus}{$\log N(\text{\hi} + \text{\htwo})$\xspace}
\newcommand{\lognhi}{$\log N(\text{\hi})$\xspace}
\newcommand{\logalam}{$\log A_\lambda$\xspace}
\newcommand{\ainh}{$A_I/N(\text{H})$\xspace}

\shorttitle{UV Extinction as a More Fundamental Dust Measure}
\shortauthors{Butler \& Salim}

\begin{document}

\title{UV Extinction as a More Fundamental Measure of Dust than \ebv or \Av}

\correspondingauthor{Robert E. Butler}
\email{bbutler3@iu.edu}

\author[0000-0003-2789-3817]{Robert E. Butler}
\affiliation{Indiana University \\
727 E. Third Street \\
Bloomington, IN 47405}

\author[0000-0003-2342-7501]{Samir Salim}
\affiliation{Indiana University \\
727 E. Third Street \\
Bloomington, IN 47405}


\begin{abstract}

The gas-to-dust ratio of reddened stars in the Milky Way (MW), the Magellanic Clouds, and in general is usually expressed as a linear relation between the hydrogen column density, \nh, and the reddening, \ebv, or extinction in the $V$ band (\Av). If the extinction curve was truly universal, the strength of the relationship and the linearity would naturally be maintained for extinction at any wavelength, and also for \nh vs. \ebv. However, extinction curves vary within the Milky Way, and there is no reason why, except by chance, either \ebv or \Av would be the most physical measure of dust column density. In this paper, we utilize for the first time full extinction curves to 41 MW sightlines and find that the scatter between \nh and extinction is minimized---and the relation becomes linear---for extinction at $2900 \pm 160$ \AA. Scatter and nonlinearity increase at longer wavelengths and are especially large for near-IR extinction. We conclude that near-UV extinction is a superior measure of the dust column density for MW dust. We provide new, non-linear gas-to-dust relations for various dust tracers. We also find that the very large discrepancy between MW and SMC gas-to-dust ratios of 0.9 dex in \nhebv is reduced to 0.7 dex for far-UV extinction, which matches the difference in cosmic abundances of carbon between the two galaxies, and therefore confirms that \nc is the preferred measure of the gas in the gas-to-dust ratio, even though it may not be a convenient one.

\end{abstract}



\section{Introduction} \label{sec:intro}

A practical quantification of the amount of dust lying between the observer and a star or galaxy is of great importance for a wide range of Galactic and extragalactic studies, including efforts to model galaxy evolution \citep{Galliano2018,Salim2020}. Ideally, the dust should be measured as the dust column density (\nd), and consequently, the gas-to-dust ratio should be expressed as \nh/\nd (or equivalently $m_\text{H}/m_\text{d}$), where \nh represents hydrogen column density between the observer and the emitter. However, measuring dust column density is not straightforward. In situations where all of the dust lies in front of the reddened object, one can use far-infrared emission to obtain dust column density. IR maps are the basis for the MW reddening maps of \citet{Schlegel1998}, which are widely used to correct (``deredden") the fluxes of extragalactic objects. However, when the dust is situated both in front of \textit{and} behind a reddened star in the Galaxy, or when a stellar population is mixed with dust in an external galaxy (galaxy attenuation), IR emission overestimates the dust column. We must then use other ways to estimate it.

Traditionally, it is the reddening (\ebv, i.e., the difference in extinction between $B$ and $V$ bands $A_B-A_V$) which is most commonly used as a measure of dust. Consequently, \nhebv is used as a measure of the gas-to-dust ratio \citep[e.g.,][]{Savage1977,Bohlin1978}. However, there is no reason why \ebv would be a better measure of column density---and, by extension, better correlate with \nh---than \Av \citep[e.g.,][]{Zhu2017}, or extinction at some other wavelength (\alam). If the extinction curve was universal at all wavelengths, i.e., if \alam normalized at some wavelength was the same for all stars (\alam/\Av = const.), the question of the most fundamental measure of dust would be moot. \alam would simply be proportional to \ebv, with the same coefficient of proportionality for all stars. Equivalently, this is to say that the \Rv parameter is constant, because by definition $A_V = R_V E(B-V)$. In reality, \Rv---and therefore the shape of the extinction curve---can vary significantly from one sightline to another, especially for high-opacity regions in the Milky Way \citep[$2<A_V<6$; e.g.,][]{CCM,Valencic2004}. 
The diversity of extinction curves implies that if \nh is best correlated with some $A_{\lambda,\text{best}}$, it will \textit{not} be so for other \alam. 

Currently, it is unknown which extinction (\alam) best correlates with \nh and is therefore the best measure of the dust column. Consequently, we also do not know if the correlation at this best \alam is better or worse than with \ebv. From a purely practical standpoint, \ebv is the most convenient measure, because its determination (unlike that of arbitrary \alam) does not require multiwavelength photometry extending into the near-IR (where extinction tends towards zero). 

It has been asserted by \citet{Jura1980} that there is no reason why either \ebv or \Av would be the best measure of dust. Instead, Jura proposed that the fundamentally better motivated measure of dust may be extinction in the near-IR (NIR), such as $A_{3.6\mu\text{m}}$ ($L$-band), but this proposal has not been confirmed empirically. 

To our knowledge, there has been only one study that attempted to predict the wavelength at which extinction is the best measure of dust. \citet{Kim1996} used the Kramers-Kronig approach \citep{Purcell1969}, together with dust grain distribution models for different values of \Rv, to determine the wavelength
at which the gas-to-dust ratio 
is independent from \Rv---a necessary condition for the minimization of scatter between \alam and \nh. For MW dust they find that this should happen at approximately 3600 \AA\ (near $U$ band). However, this result has not been widely accepted or even discussed in subsequent literature, and in any case the question has never been put to an empirical test. The aim of this paper is to do so. 

The ancillary goal of this paper is to try to find a unifying framework for disparate gas-to-dust ratios in the MW and SMC. The gas-to-dust ratio, expressed using the column density of hydrogen, is to first order constant within the Milky Way \citep{Bohlin1978}. 
However, the gas-to-dust ratio is $\sim$0.9 dex higher in the SMC,
at least when \ebv is used as the tracer of dust. This great discrepancy suggests that the fundamental relation (one that would not change from galaxy to galaxy) could involve the column density of some element that is more directly related to the dust, in place of hydrogen \citep[e.g.,][]{Clayton1985, Mathis1990, Draine2003, Welty2012}. The best candidate for this alternative abundance is carbon, the cosmic abundance of which is 0.6 dex lower in the SMC than in the MW. The difference in abundance goes a significant way toward explaining the difference in \nhebv gas-to-dust ratios, but is still 0.3 dex short. Here we investigate if the solution to this discrepancy lies in \ebv not being the most appropriate measure of dust.

In this work we take advantage of the fact that detailed extinction curves are available in the literature for a relatively large number of MW stars. 
We use \citet[][hereafter \ctfm]{fm07} extinction curve parameters in combination with hydrogen column densities from \citet[][hereafter \citetalias{g12}]{g12} to evaluate across wavelengths from the UV to the NIR which \alam is the most fundamental measure of dust. Section \ref{sec:data} describes these sources of data.
Section \ref{sec:results} describes how our sample was created, along with the methods used to determine the best measure of dust. The plausibility of metallicity driving the scatter in the \nh vs. \alam relation is also explored. 
In Section \ref{sec:discussion} we discuss our results in the context of prior results in the literature. 
The relation of our results to extinction and metallicity in the SMC is also discussed there. 


\section{Data and Sample} \label{sec:data}

We draw the main data for this study from two sources: \ctfm provides extinction curve parameters, and \ctgud gives column densities for hydrogen and metals. \ctfm represents one of the largest samples of full extinction curves for Milky Way sightlines obtained to date. \ctgud is the largest compilation of interstellar column density measurements from the literature, containing measurements toward over 3000 stars (most in the MW and some in the Magellanic Clouds) across $\sim$50 different species.  

\subsection{Extinction} \label{subsec:data_ext}
  
\ctfm mostly contains B and late O stars for which the authors were able to gather UV through IR data: UV spectrophotometry from the \textit{International Ultraviolet Explorer} (\textit{IUE}), \textit{UBV} photometry from various sources, and \textit{JHK} photometry from the Two-Micron All Sky Survey (2MASS). The original sample observed by \textit{IUE} was trimmed to remove stars with poorer data, unreddened stars, and unusual stars (Be stars, luminosity class I stars, and early/luminous O stars), leaving 328 stars in the main \ctfm sample. 

\ctfm introduced an ``extinction-without-standards" technique in which they determine extinction based on a best-fit model SED for the observed flux from each star. This model-based method eliminates the need for observations of unreddened comparison stars (the pair method).
The accuracy of the pair method can be affected by spectral mismatch between the target and comparison star. Furthermore, the comparison stars themselves are not entirely dust free and require an extinction correction.

\ctfm published the extinction curves in parameterized form, using a parameterization that is an expanded version of the one given in \citet{Fitzpatrick1990}. In particular, for the UV region ($\lambda<2700$ \AA), they use
\begin{equation} \label{eqn:extcurve}
    k(\lambda-V) =
    \begin{cases}
        c_1 + c_2 x + c_3 D(x,x_0,\gamma) & x \leq c_5,\\
        c_1 + c_2 x + c_3 D(x,x_0,\gamma) + c_4 (x-c_5)^2 & x > c_5
    \end{cases}    
\end{equation}
where $x \equiv \lambda^{-1}$, with $\lambda$ in \um. In \citet{Fitzpatrick1990}, $c_5$ was fixed at 5.9 \invum, whereas most values determined through the new parameterization are in that vicinity. $D(x,x_0,\gamma)$ represents the 2175 \AA\ bump, and is expressed using the Lorentzian-like function
\begin{equation} \label{eqn:D}
    D(x,x_0,\gamma) = \frac{x^2}{(x^2-x_0^2)^2 + x^2 \gamma^2}.
\end{equation}
where $x_0$ specifies the location of the bump and $\gamma$ modulates its width.

To describe the optical and near-IR portions of the extinction curve, \ctfm employ a cubic spline interpolation. 
In the optical range, interpolation is based on extinction at 3300, 4000, and 5530 \AA. In the near-IR ($\lambda>1$ \um), \ctfm assume that the extinction curve is universal (i.e., \alam/$A_{1 \mu\text{m}}$ = constant when $\lambda > 1$ \um, for all stars) and follows the generally assumed power-law functional form \citep{Rieke1985}:
\begin{equation}
    A_\lambda \propto \lambda^{-\beta_\text{NIR}}
\end{equation}
with the exponent \bnir = 1.84 \citep{Martin1990}. 
Assuming a fixed near-IR curve was justified in \ctfm due to the use of relatively shallow \textit{JHK} data from 2MASS.
\begin{figure}
    \centering
    \includegraphics[width=\linewidth]{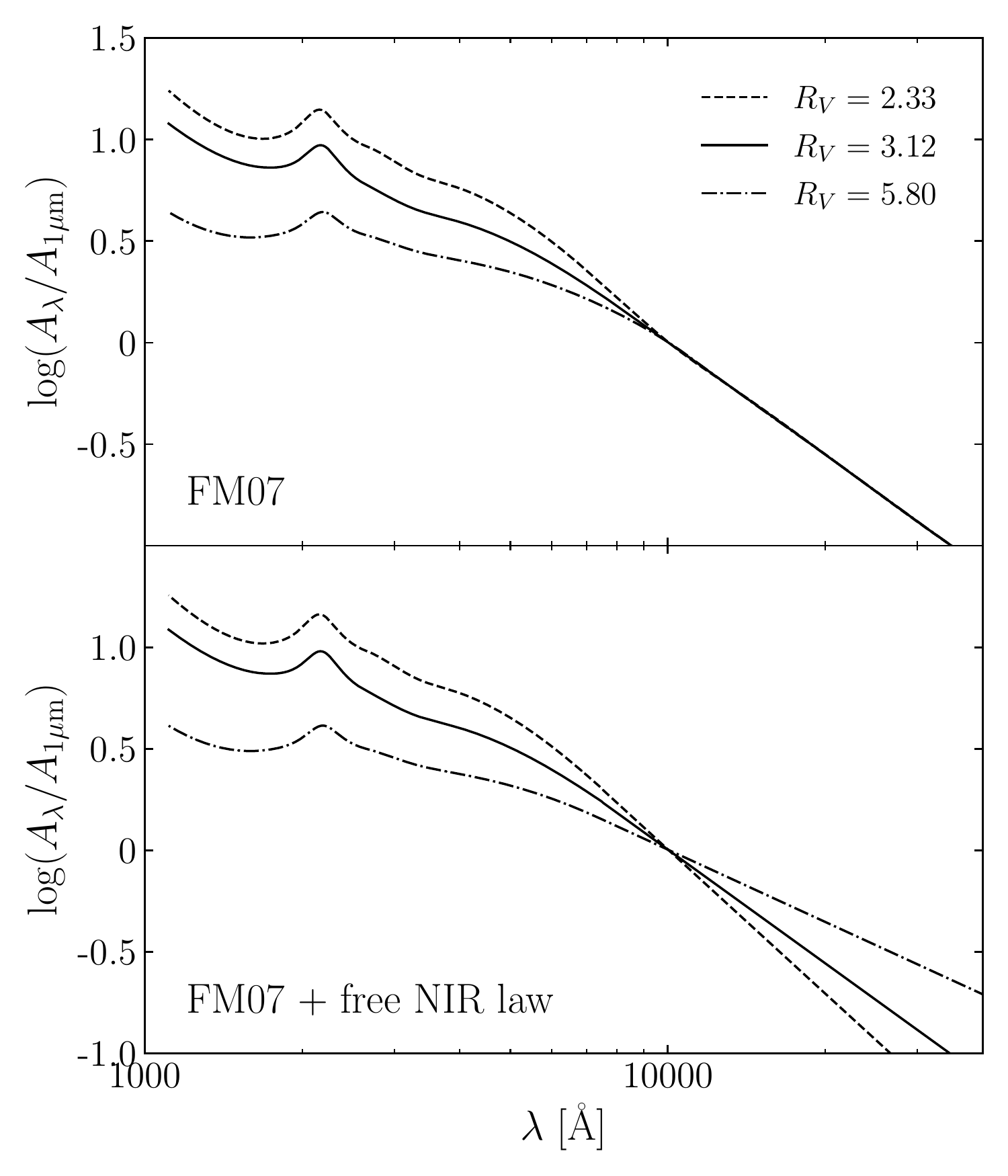}
    \caption{An illustration of the two sets of extinction curves that will be explored in this work, using stars from our sample with \Rv values of 2.33, 2.12, and 5.80. \textit{Top:} Curves produced by the \ctfm parameterization in which the near-IR extinction curve has a fixed (universal) slope. \textit{Bottom}: Same as above for $\lambda < 7500$ \AA, but allowing the near-IR extinction curve slope to be free (i.e., \Rv-dependent) as described in Section \ref{subsec:data_ext}. Note that the curves are normalized at 1 \um in order to show the fixed curve assumed in \ctfm. Also, the curves are shown in log scale, again to highlight the difference in the near-IR.}
    \label{fig:extcompare}
\end{figure}
\citet{Fitzpatrick2009} added new spectral observations from HST's Advanced Camera for Surveys High Resolution Camera to investigate the potential for flexibility in the near-IR power law slope.  They demonstrate that, of the 14 stars studied, over half of the NIR power-law fits are significantly improved by allowing \bnir to be a free parameter. They show specifically that the use of fixed $\beta_\text{NIR}=1.84$ from \citet{Martin1990} is not a good fit for large-\Rv sightlines. 
Furthermore, the slope appears to be correlated with \Rv.
Considering that there is still some debate in the literature concerning the nature of IR extinction, we produce two sets of extinction curves. One follows \ctfm, with a universal extinction curve at $\lambda>1$ \um. The other follows \ctfm up to 7500 \AA, but subsequently follows a power-law parameterization with a variable exponent \bnir. The wavelength at which the regime changes is based on the analysis in \citet{Fitzpatrick2009}. The exponent for each sightline is determined according to the relation in \citet{Salim2020} constructed from the data reported in Table 3 of \citet{Fitzpatrick2009}:
\begin{equation} \label{eqn:betacalc}
    \beta_\text{NIR} = -4.20 + 4.59 \left(\frac{A_B}{A_V}\right),
\end{equation}
where $A_B/A_V \equiv 1 + 1/R_V$. In other words, the near-IR exponent inversely depends on \Rv.
The two sets of extinction laws are illustrated in Figure \ref{fig:extcompare} for three stars from the \ctfm sample with a range of \Rv values. Note that the extinction curves with non-universal \bnir show no break in slope around 7500 \AA, even though this transition is not smooth by construction; they appear more natural. 

We use extinction curves from \citet{Gordon2003} for sightlines to SMC stars.
The authors fit reddened stars from both the SMC (wing and bar sample stars) and the LMC (LMC2 supershell and normal samples) to the parameterization in \citet{Fitzpatrick1990}. 

\subsection{Column densities} \label{subsec:data_coldens}

To obtain hydrogen column density measurements, we use the database of literature values for 3008 sightlines to stars compiled in \ctgud. \ctgud contains two tables: one with all of the measurements found in the literature, and a second with only the most recent measurements for each sightline. We use the latter. \ctgud reports \nhi and \nhtwo, along with an \nh entry, meaning \nhplus. It is notable that these \nhplus values sometimes differ somewhat from the combination of individual density values ($=N(\text{\hi}) + 2N(\text{\htwo})$). This is due to the disparate sources from which information was obtained, and the fact that individual \nhtwo and \nhi measurements have sometimes been reported in the literature more recently than \nh for the same sightline. For relative consistency, we use the \nh column as opposed to the summed individual values, and refer to it as \nhplus or \nh interchangeably.

Some column density values do not have associated errors, in which case we adopt the mean of the log-errors on the density measurements with reported errors.
The notable exception is the sightline toward HD 149757 ($\zeta$ Oph), which had an error value on \nh that was over an order of magnitude smaller than all the others, and therefore appears erroneous. We reset the error value for this star to the average (as if it did not originally have one). 

\ctgud includes column density measurements for a great many species, of which we use \nfeii, \noi, \novi, \nmgii, and \ncaii.
We pull updated hydrogen column densities for SMC stars from \citet{Welty2012}\footnote{Obtained from the full data table at \url{https://astro.uchicago.edu/~dwelty/mcoptuv.html}.}, which was published after the \ctgud compilation. 

The lines used to determine column densities for the species in question lie in the UV (except for the \caii H and K lines), necessitating the use of data from space-based instruments such as \textit{IUE}, \textit{FUSE}, STIS on \textit{HST}, etc.

\subsection{Sample selection} \label{subsec:sample}

We matched the 328 stars with extinction curves from \ctfm to the list of 3008 stars with column density measurements from \ctgud by matching the stars by name. The \ctfm stars are mostly $V<10$ mag, with a few slightly fainter; this means the stars all have names from common catalogs. 
Ninety-six stars are common to \ctfm and \ctgud, of which 70 stars have \nhi measurements, 54 have \nhtwo measurements, and 50 have \nhplus reported. 
For some stars in the \ctfm sample, \Rv was assumed to be 3.1 because of a lack of IR photometry. For others, the $O_1$ spline point was not calculated due to missing $U$-band photometry. We omit stars which fit within either category. This leaves us with 41 sightlines in the \nh sample and 55 in the \nhi-only sample.

Pertaining to the analysis of metallicity effects in Section \ref{subsec:metals}, of the 41 stars in the \nhplus sample, 28, 27, 9, 21, and 23 stars have \feii, \oi, \ovi, \mgii, and \caii, respectively. 

For the SMC analysis in Section \ref{subsec:disc-GDRSMC}, we first matched the five reddened SMC stars from \citet{Gordon2003}---four bar sample stars and one from the wing sample---with the full \ctgud list (again using names, as they all have AzV designations). Of the metal species of interest, only \caii had column densities reported (for all but one star, AzV 23). \citet{Welty2012} includes \nh measurements for all five stars. Notably, for AzV 23, AzV 214, and AzV 398 (all from bar sample), \citet{Welty2012} determined that there is no significant contribution to \nh from \nhtwo.

\begin{figure*}[t!]
    \centering
    \includegraphics[width=\textwidth,height=\textheight,keepaspectratio]{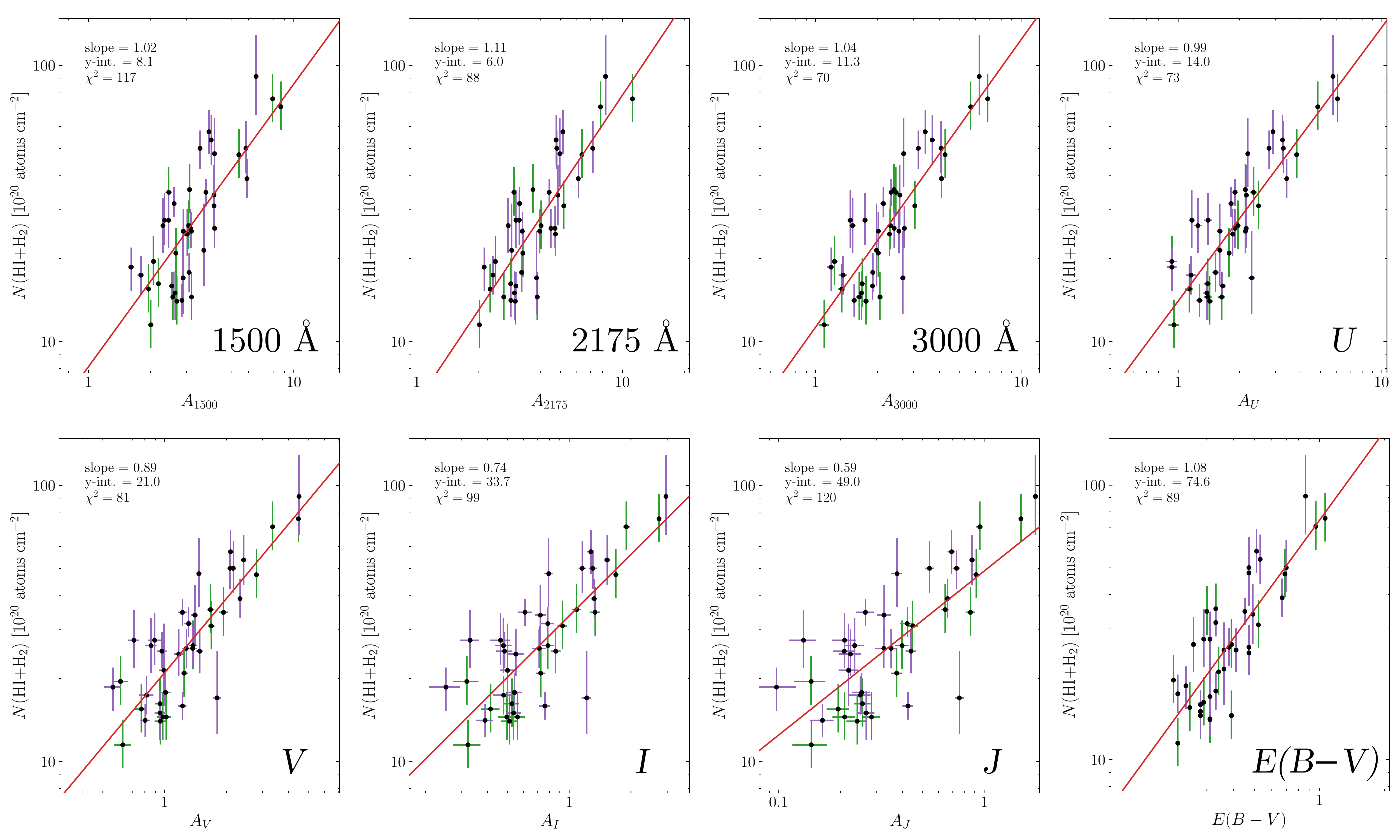}
    \caption{Relationship between total hydrogen column density and extinction at 7 selected wavelengths, along with the standard \nhplus vs. \ebv for comparison. Where the name of a photometric band is used, $\lambda$ is the standard central wavelength of the filter. Points with green errorbars are those without \nhplus errors reported in \ctgud, treated as described in Section \ref{subsec:data_coldens}. Points with purple errorbars had errors reported. In each plot, the two relevant parameters corresponding to Equation \ref{eqn:fit} are reported in the upper left. Values for $y$-intercepts are reported in units consistent with the plot axis. \chisq values from the fits to Equation \ref{eqn:fit} are also reported. The fit is best in the 3000 \AA\ panel. There are 39 degrees of freedom for all fits.} 
    \label{fig:panel_density}
\end{figure*}


\section{Results} \label{sec:results}

The main objective of our analysis is to find the wavelength at which the hydrogen column density \nh is best correlated with the extinction \alam, which will reveal the best measure for the dust column in the Milky Way. In previous studies, either \ebv or \Av has been taken as the measure for dust simply because those measures are most widely available. Furthermore, in many cases where \Av is used, it has simply been obtained through $A_V = 3.1E(B-V)$, i.e., assuming a fixed extinction curve. However, the literature also hints that the true measure of dust may be found somewhere in the NIR \citep[e.g.,][]{Jura1980, Schlafly2016}, though this has not been verified. As our analysis covers wavelengths from the far UV to $K$ band in the NIR, we are able to assess different possibilities.

It is also noteworthy that previous studies have uniformly assumed linearity in the relation between \nh and
\ebv or \Av. In the case of diverse extinction curves (a range of \Rv), we do expect linearity with respect to the \alam that is the true measure of dust (which may or may not be \Av), but not at other wavelengths.
Instead of assuming linearity, we allow a power-law dependence of \nhi and \nhplus on \alam (and \ebv). 

\subsection{\normalfont \nhplus} \label{subsec:nhplus}

Using the \ctfm extinction curve parameters for our sample of 41 sightlines, we calculate for each reddened star the extinction (\alam) for wavelengths in the range 1111 \AA\ to 47000 \AA, pseudo-continuously sampling 1000 wavelengths from the range logarithmically. Extinctions are calculated using a modified version of the \texttt{fm07} function from the \texttt{extinction} package in Python \citep{extinction.py}. We adjust the function such that it can use arbitrary curve parameters and any \Rv. Uncertainty on the resultant \alam values is calculated by re-running the script with modified \Rv values, once using $R_V + \sigma_{R_V}$, and again using $R_V - \sigma_{R_V}$ (where $\sigma_{R_V}$ is provided in \ctfm). 
The resulting range of \alam for each star should be a good approximation of the uncertainty, since the error is dominated by $\sigma_{R_V}$. After the first run, we also calculated a second set of extinction curves, this time with \Rv-dependent \bnir redward of 7500 \AA. 

Figure \ref{fig:panel_density} shows a selection of wavelengths for which \nhplus is plotted against \alam: the UV wavelengths 1500 \AA, 2175 \AA, 3000 \AA, along with the approximate central wavelengths of the standard photometric bands $U$ (3700 \AA), $V$ (5500 \AA), $I$ (8000 \AA), and $J$ (1.2 $\mu$m). The final panel shows \ebv in place of extinction at a specific wavelength. Values for $A_I$ and $A_J$ are based on extinction curves with free \bnir. Also shown are best power-law fits calculated using the least-squares method: 
\begin{equation} \label{eqn:fit}
    \log N(\text{\hi} + \text{\htwo}) = a \log A_\lambda + b,
\end{equation}
where $a = 1$ would correspond to a linear relation between \nhplus and \alam (in that case, $b$ gives the gas-to-dust ratio assuming \alam as the dust measure).  
Errors on \nhplus from \ctgud are taken into account in the fit. We calculate a \chisq value for each individual fit, which is to first order proportional to the scatter of the residuals around the best fit.

Following the original formulation from \citet{Bohlin1978}, most studies assume a linear relation, typically with $A_\lambda = A_V$, or using \ebv as the independent parameter. 
From Figure \ref{fig:panel_density} we conclude that neither option is optimal. The scatter (\chisq) is smaller in the UV and blue optical than at longer wavelengths. Also, the relation is closer to linear in the UV and blue optical than at longer wavelengths (including near-IR). The relation against \ebv is similar in quality to the relation in $V$ band and likewise is not quite linear. The inferiority of near-IR extinction as a measure of dust would also hold using a fixed NIR law, since it already sets in at $A_I$ ($\sim$8000 \AA). 

\begin{figure}[t]
    \centering
    \includegraphics[width=\linewidth]{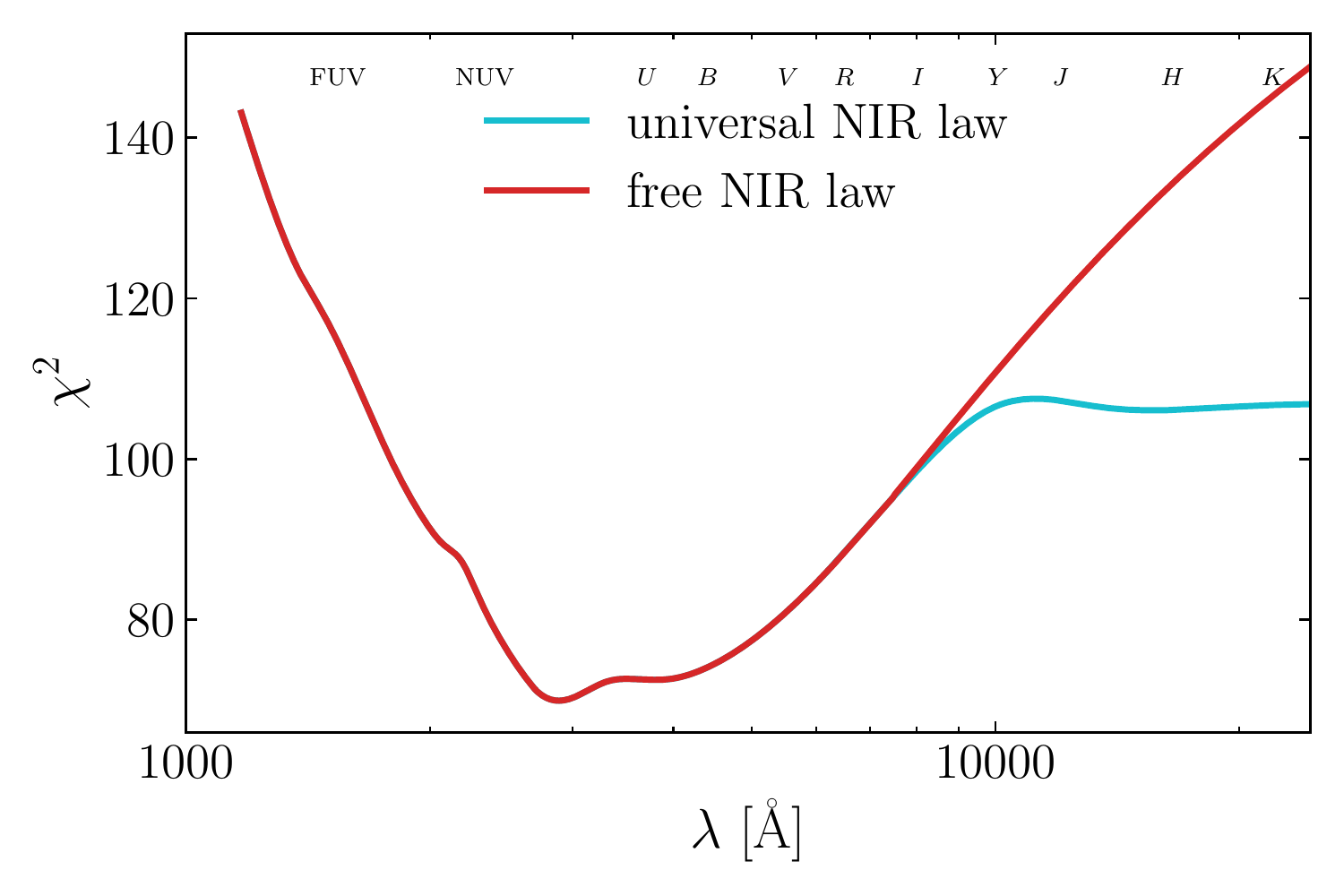}
    \caption{The sum of normalized quadratic residuals (\chisq) between \lognhplus and \logalam as a function of wavelength. The \chisq value comes from a fit to Equation \ref{eqn:fit}. There are 39 degrees of freedom. The cyan curve represents fits done using \alam values calculated using the original, unadjusted \ctfm extinction curve parameterization. The red curve is the same at $\lambda<7500$ \AA, but above that wavelength, \alam is adjusted from the \ctfm-calculated values using an \Rv-dependend near-IR extinction curve slope (\bnir; see Section \ref{subsec:data_ext}). The relation between \nhplus and extinction is tightest for extinction around 2900 \AA. Letters representing common photometric bands are centered on their central wavelengths.}
    \label{fig:chisq}
\end{figure}

\begin{figure}[t]
    \centering
    \includegraphics[width=\linewidth]{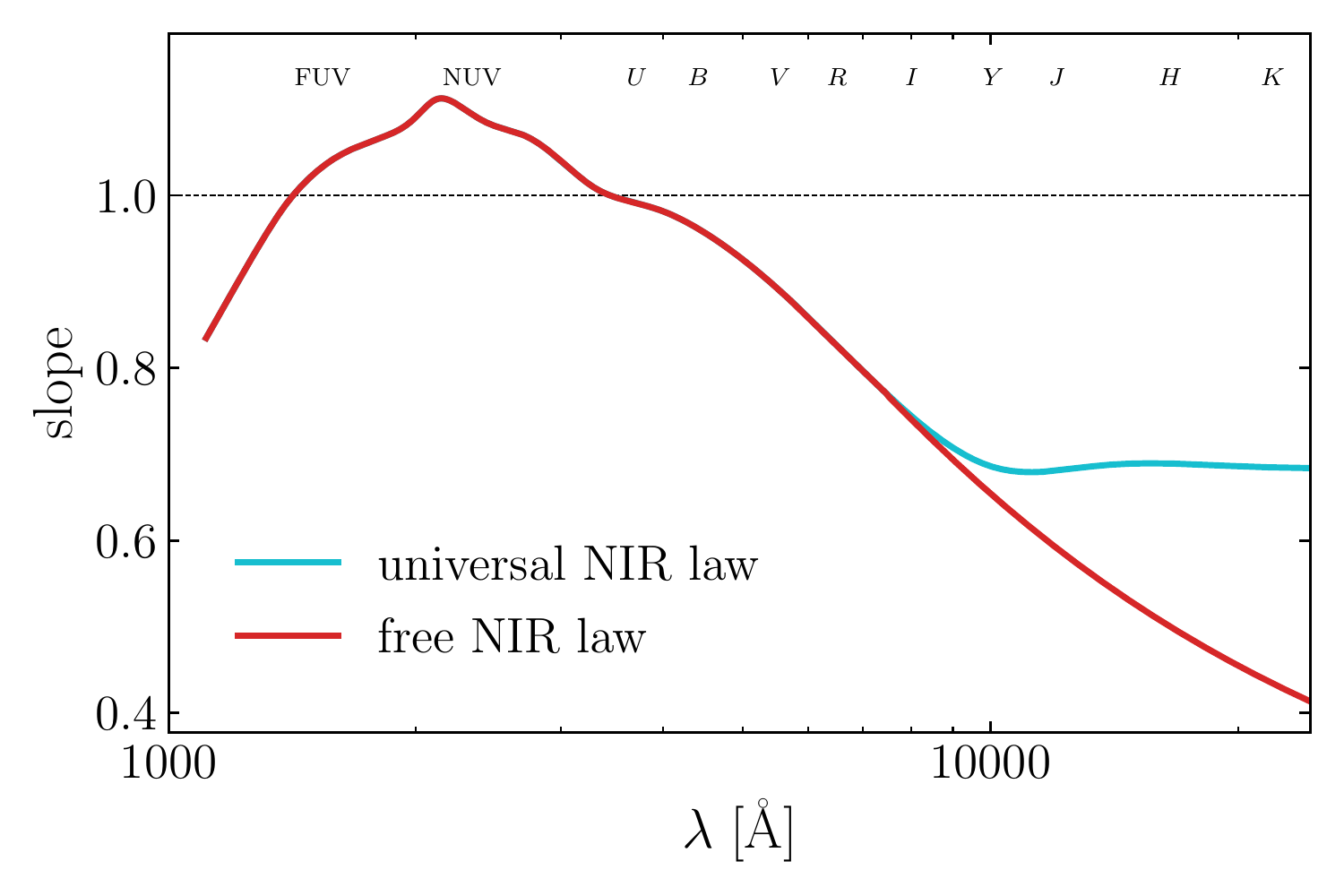}
    \caption{Slope ($a$ from Equation \ref{eqn:fit}) between \lognhplus and \logalam as a function of wavelength. The cyan curve represents fits done using \alam values calculated using the original, unadjusted \ctfm extinction curve parameterization. 
    The dotted line represents a slope of 1 (a linear relationship between total hydrogen column density and extinction). The relation between \nhplus and extinction is exactly linear for extinction around 3000 \AA.}
    \label{fig:slope}
\end{figure}

In Figures \ref{fig:chisq} and \ref{fig:slope}, we 
look at the \chisq and the slope of the best fit 
at continuous wavelengths instead of wavelengths corresponding to specific bands. Results are shown for both the universal (cyan) and free (red) NIR curves.
We show in Figure \ref{fig:chisq} that there is a wavelength region where the scatter is minimized lying in the near-UV and blue optical. Specifically, the correlation is best at $\lambda \approx 2900$ \AA, with $\chi^2_\text{red}=1.8$. This lends itself to the notion that total hydrogen column density is most closely related to dust extinction at or around that wavelength, and therefore that \Abest is the best measure of the dust column density in the Milky Way. 

To derive the error on the best-fit wavelength, we perform a bootstrapping procedure by taking 1000 random samples 
with replacement. We fit a Gaussian to the histogram of wavelengths where \chisq was lowest, restricted to the 2000 to 4000 \AA\ regime, since the distribution had high-wavelength outliers. The standard deviation of that Gaussian,
and therefore the error on the best-fit wavelength, is 160 \AA. The best-fit wavelength is therefore relatively well-constrained by our data.

It is commonly assumed that the slope of the \nh vs. \ebv relation is 1, and therefore that the slope is also 1 for \nh vs. \Av if the extinction curve does not change from star to star. Our analysis casts these assumptions in a new light. For the \ebv relation, we find the slope to be 1.08 (with $\chi^2_\text{red}=2.3$), implying possible nonlinearity, but in general agreement with the oft-assumed 1. For \Av, we find the slope to be 0.89 ($\chi^2_\text{red}=2.1$), again close to linear but not exactly so. Using a similar bootstrapping method to the one described above, we find the wavelength where the relation is exactly linear to be in the region of 3000 \AA, with error 300 \AA. This value agrees within errors with the $2900 \pm 160$ \AA\ which we find for the best fit. 
The agreement confirms that the relation \textit{is} linear between \nh and the best dust measure---which is \Abest instead of \Av or \ebv. 

For reference, the best fits are as follows:
\begin{align}
    \log N(\text{\hi} + \text{\htwo}) &= 1.05 \log A_{2900} + 21.03 \label{eqn:fit_Abest}\\ 
     &= 0.89 \log A_V + 21.32 \label{eqn:fit_Av}\\ 
     &= 1.08 \log E(B-V) + 21.87 \label{eqn:fit_ebv}
\end{align}
where the free coefficient $b$ represents the logarithm of hydrogen column density at \Abest, \Av, or \ebv of 1.
The error on \nh determined from these expressions is $\sim$0.03 dex (7\%), and is obtained by taking the typical error in \nh measurements of 0.1 dex, scaling it by \redchisq ($\sim$2), and dividing it by the square root of the sample size, resulting in the error of the mean. When compared at \ebv (or \Av) values appropriate for the respective samples, the \nh from our expressions is 10\% higher than the value from \citet{Bohlin1978}, 6\% higher than the value from \ctgud, and 8\% lower than the value derived in \citet{Zhu2017} from their analysis of the whole \citet{Anders1989} sample. These differences are well within the range of statistical error. The advantage of using our nonlinear expressions is that they acknowledge that gas-to-dust ratio is not exactly a constant, especially when \ebv or \Av is used to measure the dust.


The sightlines used in this study span the reddening range $0.21<E(B-V)<1.06$, with a median value of 0.36 for the \nhplus sample and 0.33 for the \nhi sample.
At $E(B-V) \approx 0.08$, there is a well-known break in the slope of the \nh vs. \alam relation \citep[e.g.,][]{Savage1977, Bohlin1978, Liszt2014b}. This can be attributed to \htwo formation on the surface of dust grains \citep{Hollenbach1971}, which only begins in regions characterized by sufficiently large reddening values (i.e., denser regions). Because we do not probe below $E(B-V)=0.2$, we are not sensitive to this change in slope and do not discuss it further. 

\begin{figure}[t]
    \centering
    \includegraphics[width=\linewidth]{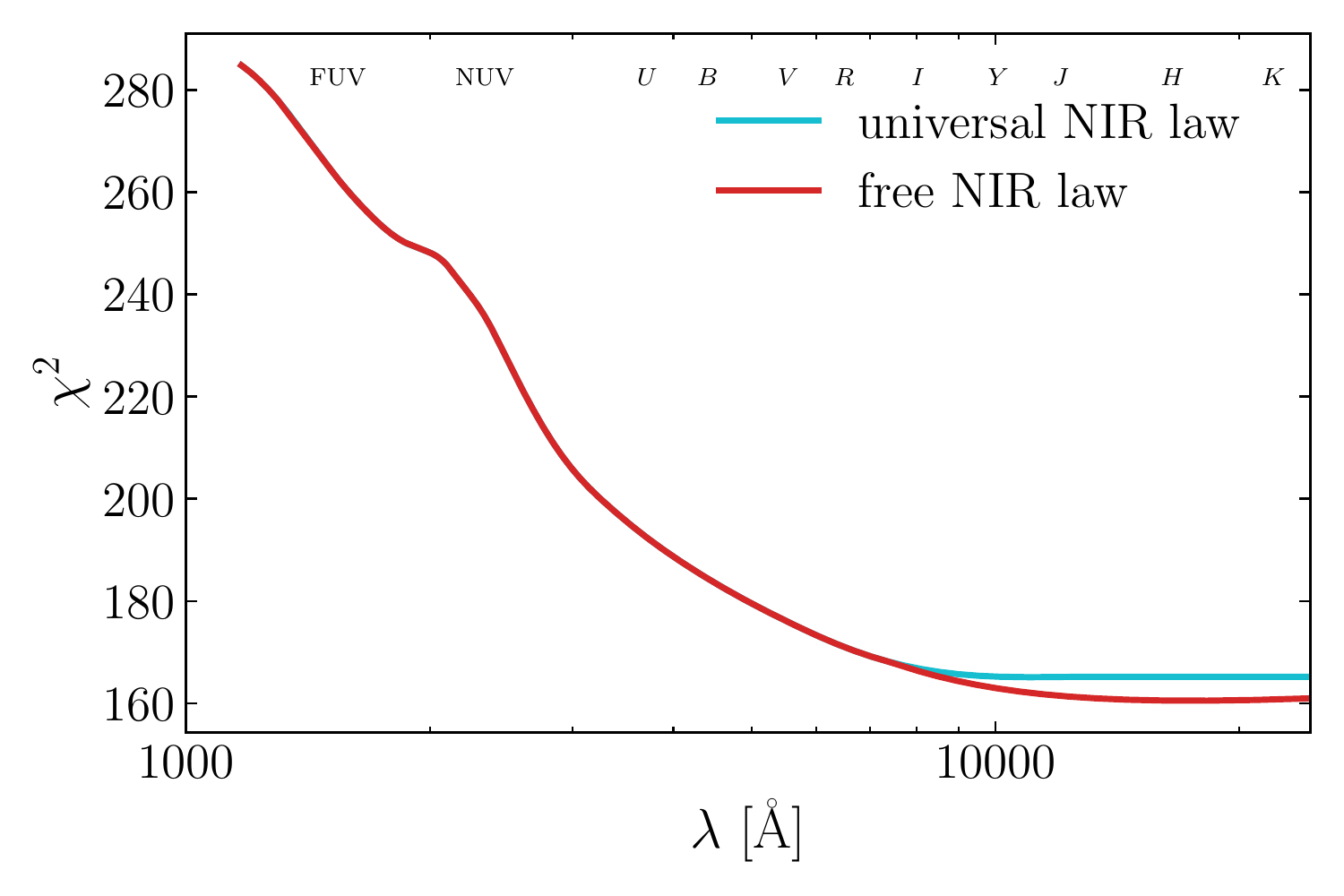}
    \caption{
    The sum of normalized quadratic residuals (\chisq) between \lognhi and \logalam as a function of wavelength. The \chisq value comes from a fit to Equation \ref{eqn:fit}. There are 53 degrees of freedom. The cyan curve represents fits done using \alam values calculated using the original, unadjusted \ctfm extinction curve parameterization. 
    The relation between \nhi and extinction is tightest for extinction around 1.7 \um, but is much less tight than for \nhplus.}
    \label{fig:hichisq}
\end{figure}

\subsection{{\normalfont \nhi} only} \label{subsec:nhi}

We carry out the same analysis as above, but using \nhi instead of \nhplus. Since \nhi measurements along Galactic sightlines are more abundant in the literature than those for \nhtwo, we are able to include 55 sightlines here.
It is well-established that the overall correlation with \ebv and \Av is stronger when using \nhplus \citep[e.g.,][]{Bohlin1978,Welty2012}. In comparing Figure \ref{fig:hichisq}, which shows \chisq values for the fit between \lognhi and \logalam, with the corresponding Figure \ref{fig:chisq} for \nhplus, one can see that we reproduce the poorer fits. At $V$, \nhplus produces $\chi^2_\text{red} \approx 2.1$, whereas \nhi produces $\chi^2_\text{red} \approx 3.4$. The $E(B-V)$ fit is actually significantly worse for \nhi, with $\chi^2_\text{red} \approx 4.6$ as opposed to $\chi^2_\text{red} \approx 2.3$ for \nhplus.

\begin{figure}[t]
    \centering
    \includegraphics[width=\linewidth]{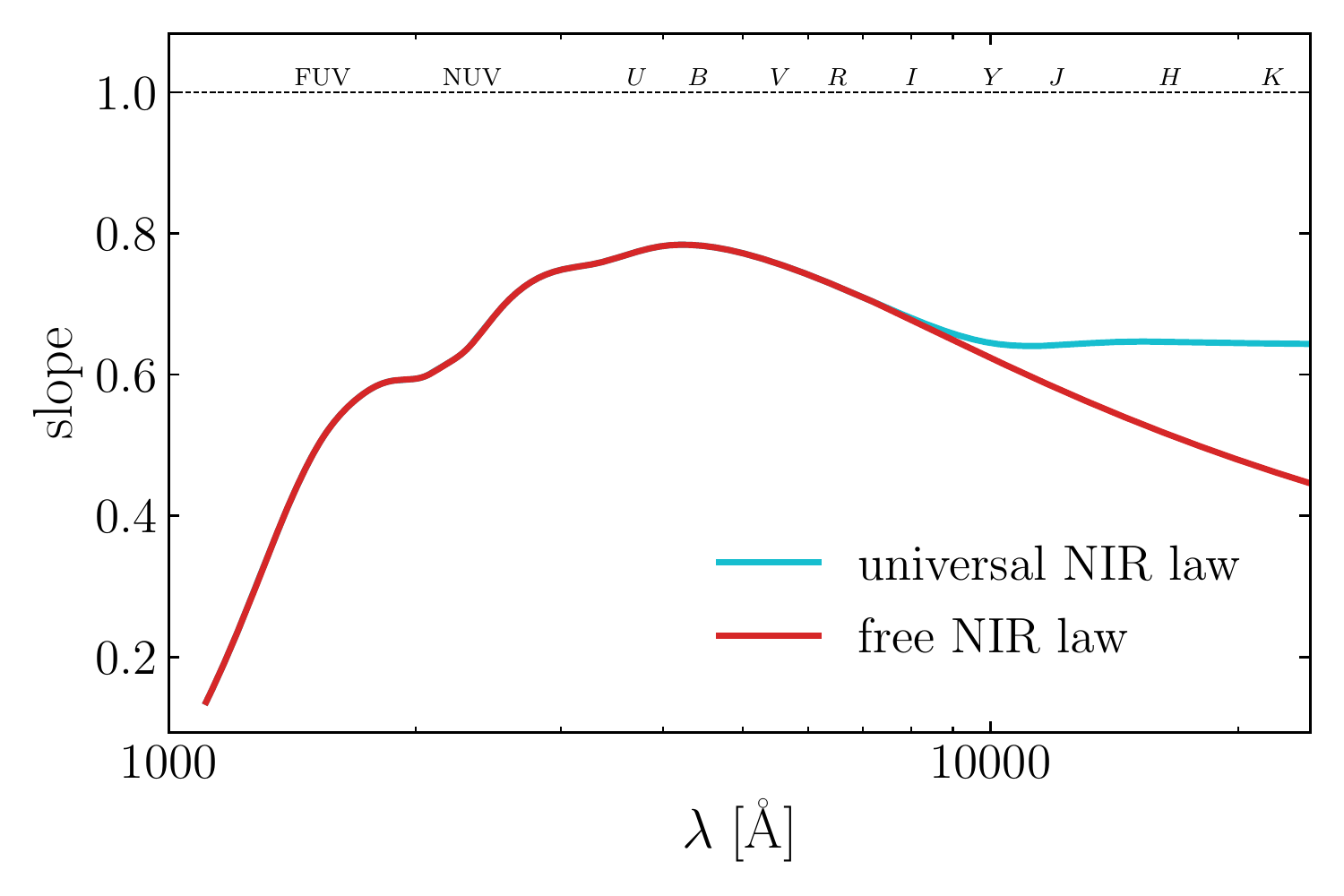} 
    \caption{Slope ($a$ from Equation \ref{eqn:fit}) between \lognhi and \logalam as a function of wavelength. The cyan curve represents fits done using \alam values calculated using the original, unadjusted \ctfm extinction curve parameterization. 
    The dotted line represents a slope of 1 (a linear relationship between total hydrogen column density and extinction). The relation between \nhi and extinction does not approach linearity.}
    \label{fig:hislope}
\end{figure}

It is noteworthy that for \nhi the best fit does not lie in the visible range; rather, the fits gradually improve toward in the NIR, stabilizing beyond about 1 $\mu$m (see Figure \ref{fig:hichisq}). We concluded in Section \ref{subsec:nhplus} that a minimum value exists at $\lambda \approx 2900$ \AA, but \nhi does not seem to be connected to the dust in the same way, as expected given that \htwo is known to form on the surface of grains. 
In Figure \ref{fig:hislope}, we see that the relation never approaches linearity, again contrasting with the \nhplus case. This behavior can be attributed to the fact that we are not counting all of the hydrogen.

\begin{figure*}[t!]
    \centering
    \hspace*{-1.1cm} \includegraphics[width=1.13\linewidth]{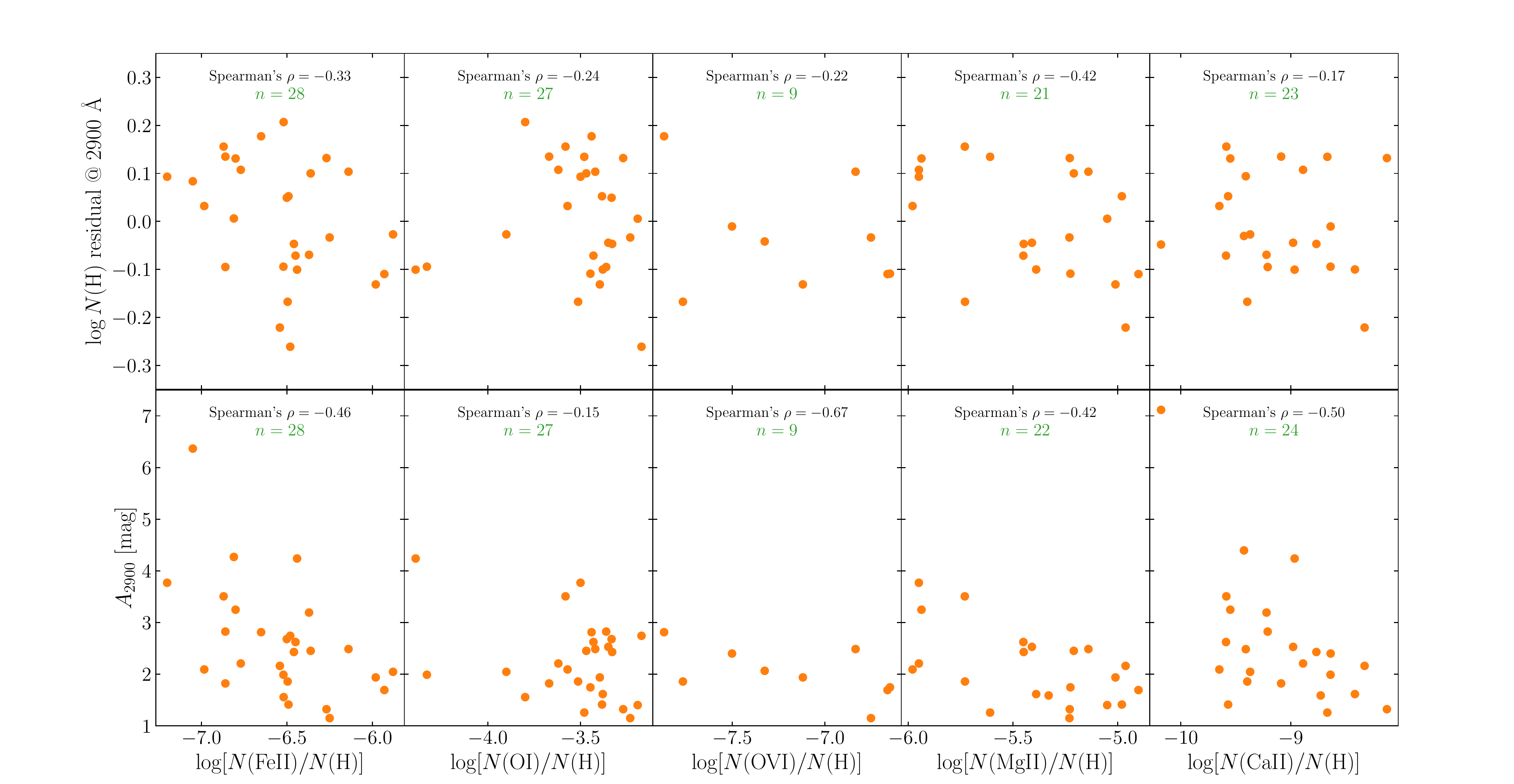}
    \caption{Metallicity, gas-to-dust ratio, and depletion of metals onto dust grains. 
    \textit{Top:} The dependence of the residuals in the \nhplus vs \alam relation (i.e. gas-to-dust ratio) on metallicity, which is represented by the column density of five different metals normalized by the total hydrogen density. 
    \textit{Bottom:} Extinction at $A_{2900}$ (our preferred measure of dust) vs. the column density of metals.
    The Spearman correlation coefficient is given in each panel, along with the total number of sightlines with a column density for the given species reported in \ctgud.
    While the correlations are mostly quite weak and some are not statistically significant, they are all in the negative direction, which is what we expect if the higher \textit{cosmic} abundance (which the \textit{observed} abundances shown reflect only partially) are related to higher gas-to-dust ratio (upper panels) and if the higher \textit{observed} abundances correspond to less depleted sightlines (i.e., lower extinction; lower panels).}
    \label{fig:metals}
\end{figure*}

\subsection{Metallicity effects on the gas-to-dust ratio} \label{subsec:metals}

Though \nhplus is strongly correlated with \alam, especially in the near-UV/blue optical, \redchisq is not 1, indicating excess scatter not described by the errors. Excess scatter has been a consistent feature of such relations in the literature with \ebv and \Av used to represent the dust, but even at \Abest (which we claim is the true best dust measure, and therefore any scatter due to differences in the extinction curve is minimized; see Section \ref{sec:discussion}), the errors on \nhplus do not encapsulate the scatter. 
A leading possibility for the residual scatter is variation in metallicity \citep[e.g.,][]{Welty2012, Remy-Ruyer2014, Kahre2018}. If one accounts for differing metallicity between sightlines, it may be possible to bring the \redchisq nearer to unity. 

We make use of 
common metallicity indicators (\feii, \oi, \ovi, \mgii, and \caii) in the \ctgud compilation to assess metallicity effects among the sightlines toward our stars.
To test whether metallicity is in fact correlated with residuals (thus driving the scatter), we plot the residual at $\lambda=2900$ \AA\ from Section \ref{subsec:nhplus}
(essentially the gas-to-dust ratio) vs. the column density of the metal species, normalized by \nhplus.  

The results are shown in the top row of Figure \ref{fig:metals}. We also provide Spearman's rank correlation coefficient $(-1<\rho<1)$ in each panel, which assesses the correlation nonparametrically (i.e., does not assume linearity). Though most correlations are weak or nonexistent, it is significant that all values of Spearman's $\rho$ are negative, implying that a very weak correlation with metal densities is not ruled out. 

The inability of the metallicity to account for the scatter in the gas-to-dust relation, even when dust is measured as \Abest, may seem disappointing. However, it is not clear that one can use  the \textit{observed} metal abundances alone to test the hypothesis in the first place. 
In fact, many metal species are subject to depletion onto dust grains themselves, so that the gas-phase abundances we measure towards these stars do not represent the true ``cosmic" abundance along any given sightline. For Si in particular, depletion is known to essentially entirely offset the observed gas metallicity differences among the Milky Way and the Magellanic Clouds \citep{Roman-Duval2019}. In the ISM, among the species we investigate, Fe and Ca are likely substantially depleted, with the Ca depletion depending more strongly on gas density; Mg is largely undepleted in diffuse gas but depleted in denser gas; and O is not strongly depleted \citep[all inferences from][]{Mathis1990}. If depletion (rather than the intrinsic range of cosmic abundances) is indeed the principal cause of the observed gas metallicity differences between sightlines, there should exist a correlation between depletion and dust density, which may be reflected in a correlation between the observed metallicity and extinction. 
In the bottom five panels of Figure \ref{fig:metals}, we show extinction 
vs. observed metallicities. For \feii and \caii, there is a relatively clear negative correlation between \Abest and normalized column density---the more dust there is, the more the metals are depleted onto it. The correlation is somewhat weaker but still present for \mgii, which aligns again with \citet{Mathis1990}. The two O species seem to be subject to relatively little depletion, though still have weak negative 
gradients. Connecting the lower panels with those in the top, it appears that it is not possible to assess the metallicity effects on the gas-to-dust ratio within the Milky Way. 

\begin{figure}[t]
    \centering
    \includegraphics[width=\linewidth,keepaspectratio]{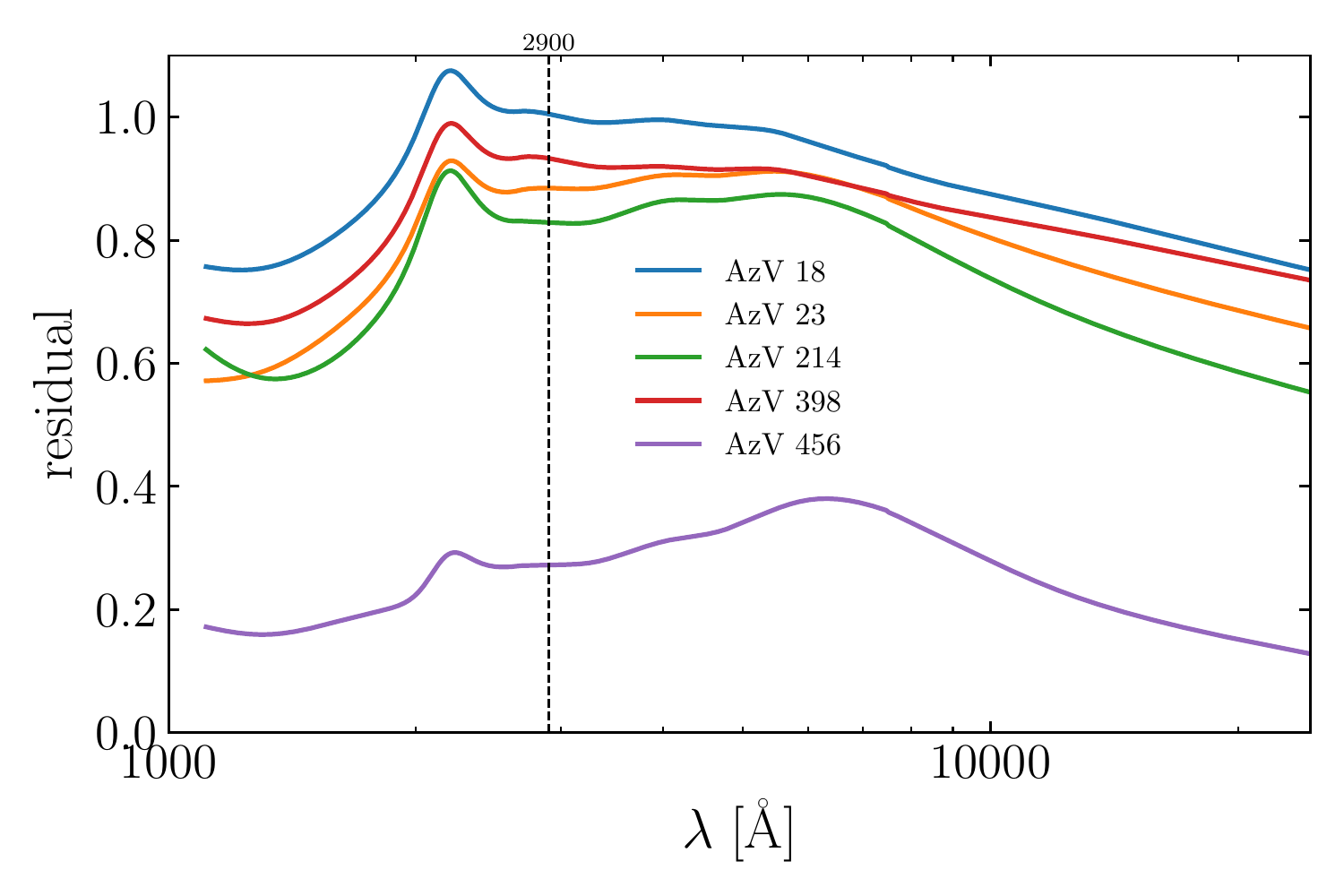}
    \caption{Residuals of \lognhplus of SMC stars from \citet{Gordon2003} with respect to \lognhplus for MW sightlines from the UV to the near-IR. The dashed vertical line represents the best-fit wavelength from Section \ref{subsec:nhplus} (=2900 \AA).}
    \label{fig:smc_resids}
\end{figure}

\subsection{SMC} \label{subsec:SMC}

So far we have focused only on the Milky Way. In the context of hydrogen column density vs. extinction/reddening, the SMC has relatively low $E(B-V)$, but high \nhplus. This would place most SMC sightlines in the upper-left region of the \nhplus vs. \ebv panel in Figure \ref{fig:panel_density}, far above the MW best-fit line. To quantify the deviation, we pull the 5 SMC stars for which \citet{Gordon2003} derived a full extinction curve, four from the bar sample, and one from the wing sample. The wing sample shows extinction features much more similar to the MW than the bar does, which \citet{Gordon1998} propose is a result of weaker star formation in the wing region. It has been suggested that the strong positive offset (for the bar sample especially) is due to the lower-metallicity environment in the SMC \citep[e.g.,][]{Welty2012, Remy-Ruyer2014}. We find this offset to be 0.9 dex when \ebv is used as the measure of dust. The question we wish to answer here is how much of the offset may be the result of not measuring the dust using a more adequate tracer.

Figure \ref{fig:smc_resids} shows the residuals of SMC sightlines with respect to the \nhplus vs \alam relations for MW sightlines carried out in Section \ref{subsec:nhplus}.
The plot confirms the large offsets in the optical region (around $V$ band), especially for the bar sample stars. However, the offsets are clearly reduced in the UV, albeit at wavelengths shorter than 3000 \AA. This lends more evidence toward our conclusion from Section \ref{subsec:nhplus} that dust is better measured using extinctions at UV wavelengths. We discuss a possible framework for bringing the MW and SMC gas-to-dust ratios into full agreement in Section \ref{subsec:disc-GDRSMC}.

\section{Discussion} \label{sec:discussion}

\subsection{A physically motivated measure of the dust column for the Milky Way} \label{subsec:disc-dustmeasure}

Through studying the correlation between \nh and \alam from the UV to the NIR, we find that the extinction at 2900 \AA\ (\Abest) is more fundamental than \ebv or \Av as a measure of dust, at least for the MW. Further, the slope of the relation is equivalent to unity around this wavelength, solidifying the result.

\begin{figure*}[t]
    \centering
    \includegraphics[width=\textwidth,height=\textheight,keepaspectratio]{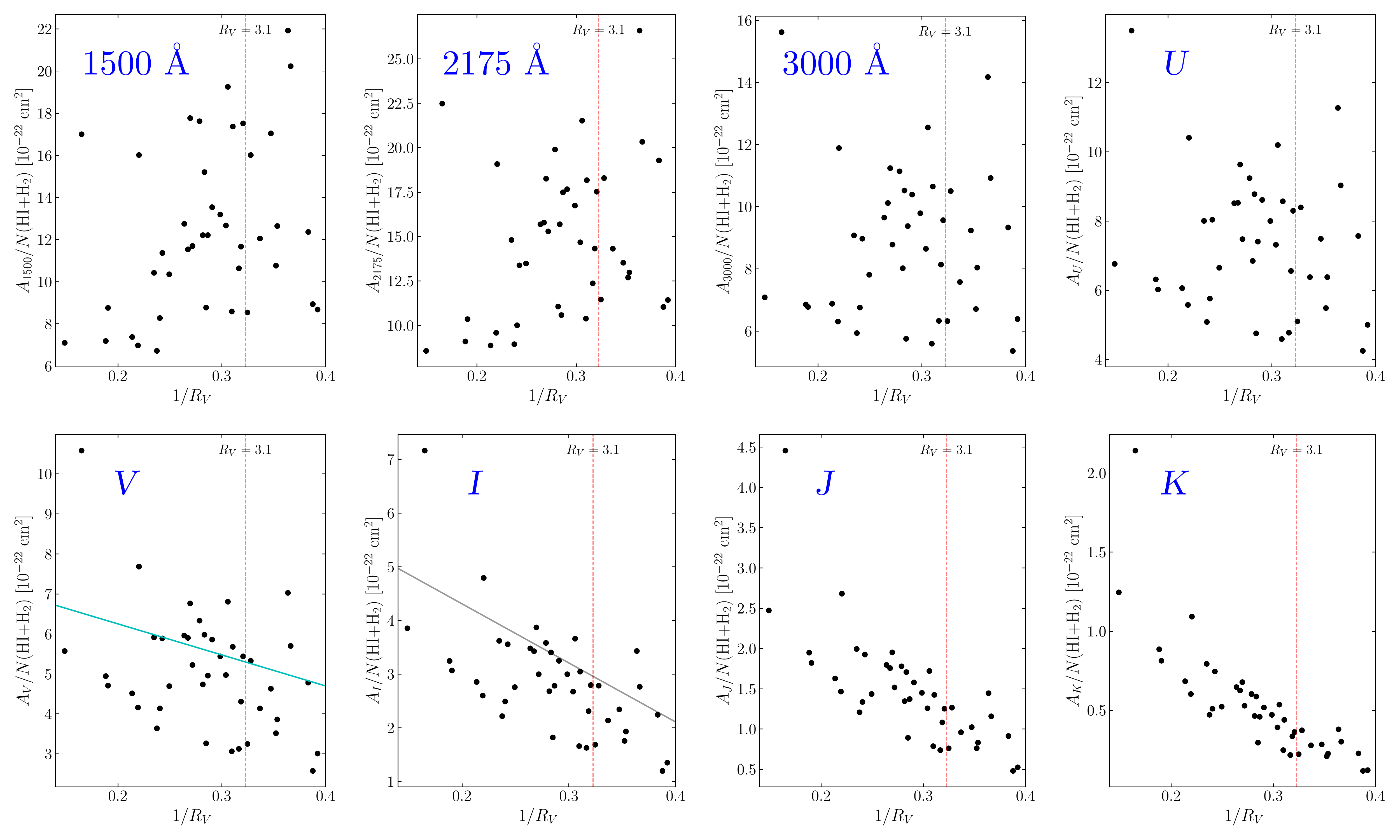}
    \caption{Dust-to-gas ratio for 41 individual sightlines vs. $1/R_V$, where the dust is represented by extinction at specific wavelengths from the UV to the NIR, as indicated by the labels. Vertical dashed line indicates where $R_V=3.1$, for convenience. Note that $y$-axis ranges vary. For the last three panels ($I$, $J$, $K$), the extinction values are derived using the variable NIR extinction curve slope (\bnir; see Section \ref{subsec:data_ext}). The cyan line on the $A_V$ panel represents predictions from \citet{Kim1996}. The $A_I$ panel is analogous to Figure 2 from \citet{Draine2003}, and the gray line represents the best fit from that plot. Dust-to-gas ratio is independent from \Rv at 3000 \AA and at $U$, confirming the results obtained earlier in this work.}
    \label{fig:axrv}
\end{figure*}


Does this result have any theoretical basis? As we pointed out in the introduction, \citet{Kim1996} mention that \avnh is not necessarily the best measure of the dust to gas mass ratio. Half a century ago, \citet{Purcell1969} suggested an alternative to \avnh based on the application of the Kramers-Kronig relations to the interstellar medium. For a single grain composition, the integral relation can be derived from the original Kramers-Kronig expressions \citep{Purcell1969, Martin1978, Bohren1983}.
The idea is to integrate the total extinction cross-section of the grains across all wavelengths,
by integration of the extinction curve \alam (or $A_w$, where $w\equiv \lambda^{-1}$):
\begin{align}  \label{eqn:kk2}
    \int_0^\infty \frac{A_w}{N(\text{H})w}\; d \ln w &= 1.086\pi^2 m_\text{H} \sum \frac{f}{s_d} \frac{M_d}{M_\text{H}}.
\end{align}
Here, $s_d$ is grain density (which depends on composition), and $M_d$ is the column mass density of dust corresponding to hydrogen column mass density $M_\text{H} = m_\text{H} N_\text{H}$.

\citet{Kim1996} plot the integrand from Equation \ref{eqn:kk2} for a combination of silicate and graphite grains, with separate curves for extinction curves with varying values of $R_V$. The authors point out that all curves cross near 2.8 \invum (3600 \AA). This point of invariance of gas-to-dust \textit{mass} ratios with $R_V$ coincides roughly with our best-\chisq wavelength of 2900 \AA\ from Section \ref{subsec:nhplus}.
While these results do not coincide precisely, their separation is not large considering the modest increase in our \chisq value from 2900 \AA\ to 3600 \AA\ (see Figure \ref{fig:chisq}). The \citet{Kim1996} result is also in general agreement with the place we find the slope of the \nh/\alam relation to be 1 ($\sim$300$0\pm 300$ \AA). Referencing Figure \ref{fig:slope}, one can see that the crossing point we find without bootstrapping is quite close to the $U$-band, which is centered very nearby 3600 \AA. 

In a similarly mass-based pursuit, \citet{Barbaro2004} select lines of sight in the solar neighborhood with anomalously high \nhi/\ebv. They find that the mass density ratio of gas to dust ($\rho_\text{H}$/$\rho_\text{d}$) is linearly related to \nhebv, which demonstrates that \ebv cannot be the fundamental measure of dust, as we also find. They also note that, for anomalous sightlines, a modification of $\rho_\text{H}$/$\rho_\text{d}$ compared with the Galactic standard is required. 

As the above analysis based on the Kramers-Kronig approach also demonstrates, if some \alam is truly the best measure of dust, we expect the gas-to-``dust" ratio (where the extinction \alam represents dust) to be independent of \Rv. 
We perform a test for \Rv dependence across all wavelengths probed in the main analysis (UV to NIR), and show a selection of wavelengths in Figure \ref{fig:axrv}. 
Figure \ref{fig:axrv} shows the wavelength regime where \alamnh is independent of \Rv: somewhere around 3000 \AA\ or $U$ band. This agrees with our result obtained using a different method (minimization of scatter between \nh and \alam).
While the switch from slight positive to slight negative correlations in the plots is subtle, it is clear that at \Av and beyond, there is a definite relationship between \alamnh and \Rv, providing more evidence for ruling out extinction at those wavelengths as a satisfactory dust measure. \citet{Kim1996} found that as \Rv increased beyond the standard 3.1, their maximum entropy solution did not use up as much material, yielding increased \avnh. These predictions are shown in cyan on the $V$ panel of Figure \ref{fig:axrv}. \citet{Draine2003} constructs an \Rv-dependent relation for \ainh, using 14 sightlines through translucent clouds from \citet{Rachford2002}. The resultant relation is shown in gray on the $I$ panel of Figure \ref{fig:axrv}. Both the \citet{Kim1996} and \citet{Draine2003} relations agree with what we find. 

\subsection{The character of the near-IR extinction curve} \label{subsec:disc-NIR}


In our analysis, we have produced the near-IR parts of the extinction curves by two different methods: (1) assuming universality beyond 1 \um (following the \ctfm parameterization that imposes it), and (2) modifying the \ctfm parameterization beyond 0.75 \um in order to allow the NIR slope to vary with \Rv, following the results from \citet{Fitzpatrick2009}. We see that our results are entirely unaffected by this assumption. 
The reason for considering NIR extinction as a more physically motivated measure lies in the notion that the NIR extinction curve may be universal \citep{Rieke1985,CCM}. A universal curve in the NIR means that the ratio of extinctions for any two wavelengths in the NIR range will be constant, even if the shape of the curve in the UV/optical is variable \citep[and, following][is correlated with \Rv]{CCM}. However, the possibility that the NIR extinction curve is universal only guarantees that any NIR extinction (e.g., $A_I$, $A_J$, $A_K$) will be an equally good (or bad) measure of the dust, but not necessarily the best measure. In other words, the question of universality in the near-IR has no bearing on which extinction measure is most closely related to dust column density.

While shedding light on the question of the character of the near-IR extinction curve is not the goal of this paper, we overview the subject briefly.
The foundational study \citet{CCM} finds that the IR data of the time was consistent with a single extinction law for $\lambda > 0.90$ $\mu$m, citing several contemporary studies \citep{Jones1980, Koornneef1983, Rieke1985, Smith1987, Whittet1992} and ultimately using the \citet{Rieke1985} curve as the basis for the \bnir used in their parameterization, with a value of 1.61. 
More recently, with deeper NIR data available, higher values have been found: 1.95 \citep{Wang2014}, 2.07 \citet{Wang2019}, and upward. 
Studies with the highest values \citep[e.g.,][]{Naoi2007, Nogueras-Lara2018} focus on extremely dense regions (like the Galactic Center and the Coalsack). Steeper slopes may correlate with the density of the regions probed, pointing to non-universality. Indeed, it may be claimed that the wealth of recent works whose results span a wide range of \bnir values themselves contribute to a variable \bnir as a conglomerate. 

As referenced in Section \ref{subsec:data_ext}, the strongest case for non-universality may come from \citet{Fitzpatrick2009}. 
The resultant values for \bnir varied from 0.9 to 2.3, and their Figure 3 reveals a correlation between near-IR slope and optical slope (and therefore \Rv) for the 14 sightlines. \citet{Zasowski2009} shows general agreement with the \citet{Fitzpatrick2009} result, citing a variance in \bnir with Galactocentric radius which seems to be consistent with higher values being found toward the Galactic Center.
Similarly, \citet{Schlafly2016} find that their extinctions derived using the standard crayon technique agree better with the \citet{Fitzpatrick2009} modification of the \ctfm curve than the \ctfm curve itself. They fit extinction curves to their reddening vector measurements by computing $dm_b/dA$, where $dA$ is a small variation in extinction about a typical APOGEE $E(B-V)=0.65$, and $m_b$ is the observed magnitude in some bandpass $b$. This is done for bandpasses in the optical ($grizy$) and the NIR ($H$, $K$, W1, W2). The resultant fit ($\chi^2=26$) for \citet{Fitzpatrick2009} curves---the only ones tested which allowed \bnir to vary---was quite a bit better than for prior curves more similar to \ctfm, the closest contender being \citet[][$\chi^2=93$]{Fitzpatrick1999}. 
We point out that Figures \ref{fig:extcompare}, \ref{fig:chisq}, \ref{fig:slope}, and \ref{fig:hislope} do seem to suggest, but not prove, that a non-universal IR law is more natural because it provides a continuation of the trends seen at shorter wavelengths.

\subsection{Dependence of the gas-to-dust ratio on metallicity and the SMC} \label{subsec:disc-GDRSMC}

In the landscape of plots showing \nh vs. \ebv, there is always quite a bit of scatter. \citet{Bohlin1978}, found this scatter to be $\sim$30\% about the mean. In later studies \citep[such as][]{Rachford2009}, the scatter remains. In all cases, the scatter is larger than can be explained by the errors. In our analysis, the scatter has been partially mitigated by substituting \Abest as the measure of dust. However, the resultant \redchisq still exceeds unity. We attempted to invoke metallicity effects to explain this in Section \ref{subsec:metals}, but as noted, the depletion of various species renders the analysis murky.


The gas-to-dust ratio is well studied in the literature, both in the Galaxy and beyond, and versus a range of other parameters. Using a variety of Galactic X-ray sources, \citet{Zhu2017} find no correlation between \nhav and hydrogen number density, Galactocentric radius, or distance from the Galactic plane, where the latter two were looked into as a way to explore potential dependence on metallicity. Our results suggest that 
probing the gas-to-dust ratio using \Av, where the residuals are also driven by differences in the extinction curve (Figure \ref{fig:axrv}), dilutes any dependence on metallicity, likely leading to the null result.
\citet{Vuong2003} also use Galactic X-ray sources to study the ratio, but in nearby dense clouds, where the ratio (using $A_J$) in the $\rho$ Oph cloud in particular is found to deviate significantly below the Galactic value. They extend this by finding that the difference between Galactic and local values for the ratio are entirely due to metallicity differences. 
\citet{Kahre2018} find a power-law relationship between the gas-to-dust ratio and gas-phase oxygen abundance measured from nebular emission in five nearby galaxies, along with a change in this relationship if \htwo is excluded from the gas tally. This result, based on \textit{different} galaxies, is not necessarily in contradiction to our result---the lack of a strong correlation between gas-to-dust ratio and metallicity \textit{within} our Galaxy, where the observed metallicity does not correspond to cosmic abundances as it does for global measurements based on \htwo regions.


\begin{figure}[t]
    \centering
    \includegraphics[width=\linewidth,keepaspectratio]{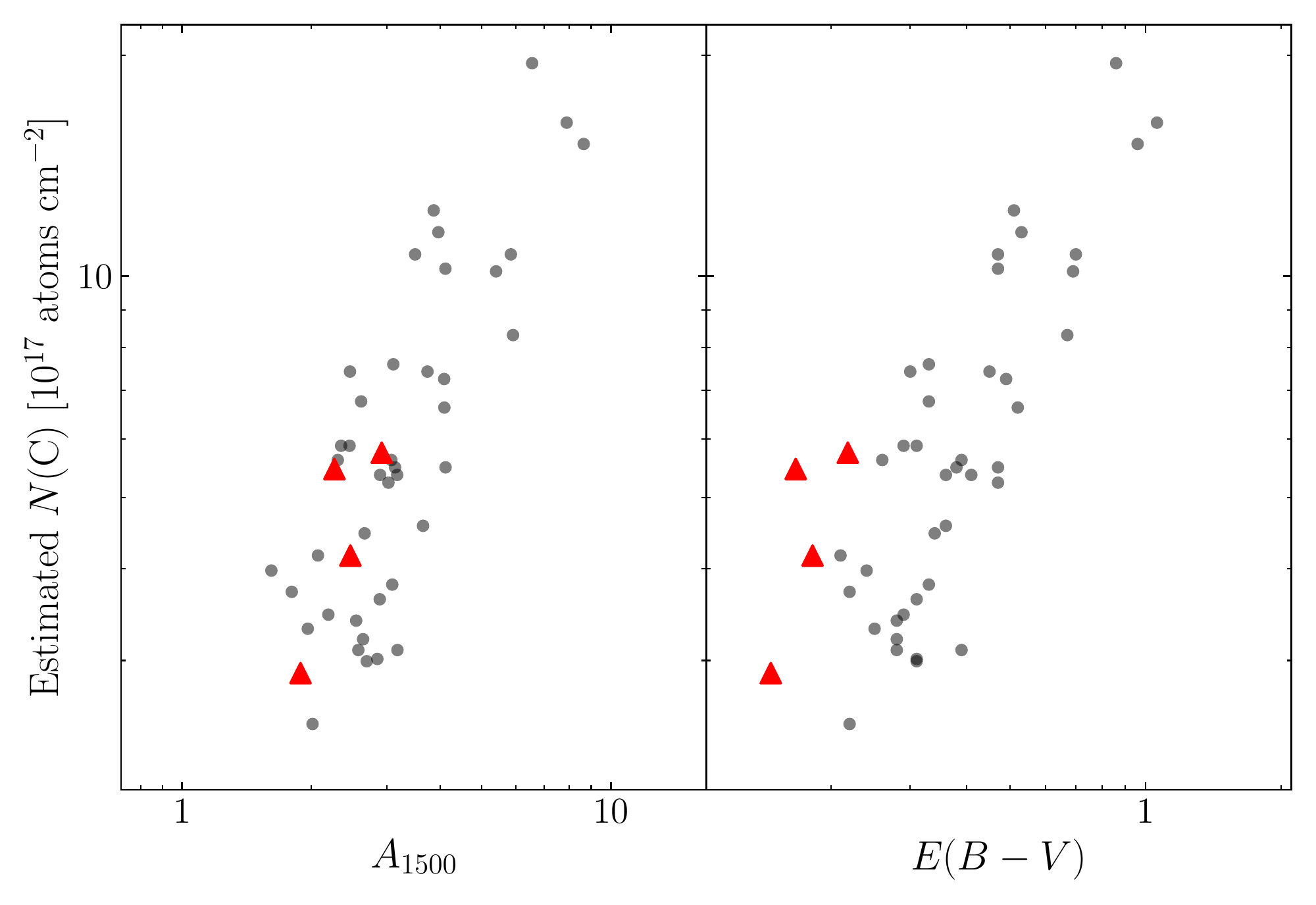}
    \caption{Estimated ``cosmic" carbon column density for the 41 MW sightlines from Section \ref{subsec:nhplus} (gray points), overplotted with 4 sightlines to SMC bar sample stars from \citet[][red triangles]{Gordon2003}, vs. extinction at 1500 \AA\ (left) and $E(B-V)$ (right). \nc is estimated through adjusting the original \nh using constant shifts derived from $12+\log(\text{C/H})$ values for the MW \citep[solar vicinity;][]{Russell1992} or SMC \citep{ToribioSanCipriano2017}.}
    \label{fig:nc}
\end{figure}
 
In Section \ref{subsec:SMC}, we found that the difference in gas-to-dust ratio between the SMC and the MW using $A_{1500}$ as the dust measure 
is smaller than using \Av or \ebv, but still persists. Interestingly, the remaining difference ($\sim$0.7 dex) matches the difference in the ``cosmic" abundances of carbon between the Milky Way and the SMC \citep{Russell1992, Clayton2000a, ToribioSanCipriano2017}, but not silicon \citep[0.3 dex;][]{Russell1992, Clayton2000a}. 
The relationship between differences in gas-to-dust ratio and differences in carbon abundance seems to be somewhat understood in the literature \citep{Clayton1985,Mathis1990,Draine2003,Welty2012}, 
but exact agreement is achieved only when dust is measured using far-UV extinction (near UV is problematic because of the UV bump; Figure \ref{fig:smc_resids}). 
We illustrate the effects of switching to \nc in Figure \ref{fig:nc}, which shows the 4 bar-sample SMC sightlines from \citet{Gordon2003} overplotted onto two of the panels from Figure \ref{fig:panel_density}. In the right panel, it is clearly shown that the difference in cosmic abundance is \textit{not} the sole factor in explaining the gas-to-dust ratio difference between the Milky Way and the SMC when one uses \ebv to measure dust. However, the left panel shows that at 1500 \AA, the difference in abundances \textit{does} account for the disparity.  


\section{Conclusions} \label{sec:conclusions}
In this work, we set out to find the most fundamental measure of dust, and consequently of the dust to gas ratio. The main conclusions from this study are as follows.
\begin{enumerate}
    \item The strongest correlation between total hydrogen column density \nhplus and extinction \alam for Milky Way sightlines is found at $\lambda \approx 2900$ \AA; furthermore, the correlation is linear at that wavelength. 
    \Abest is a superior measure of dust to \ebv and \Av.

    \item The result that \Abest is a more fundamental measure of dust supports a prediction by \citet{Kim1996} through employment of the Kramers-Kronig relation for Milky Way dust. This approach also reveals that the wavelength of the extinction value which is the most physically motivated dust measure is grain size and grain composition specific, but generally lies in the UV.
    
    \item We find no strong correlation between residuals from the \nhplus vs. \Abest relation and metal column densities. However, we show that such an assessment of the drivers of the gas-to-dust ratio is not possible using the metallicities measured in the same sightlines as the reddened stars because of the depletion of metals onto dust grains. Indeed, we do see some evidence of depletion in the form of a correlation between \Abest and $N($metal)/\nh.
    
    \item The correlations between \nhi and \alam are uniformly weaker than those for \nhplus, and are not linear for any \alam.
    
    \item The progression of the quality of the fits at longer wavelengths appears to proceed more naturally if we assume an \Rv-dependent NIR extinction curve slope (\bnir) than if we assume a fixed slope (i.e., a universal NIR curve).
    
    \item Extinction at UV wavelengths also helps reduce the gap between the gas-to-dust ratios measured for the SMC and the Milky Way. This difference is 0.9 dex when dust is measured as \Av and 0.7 dex for $A_{1500}$. The latter value agrees with the difference in cosmic abundances of carbon between the SMC and MW, lending support to the idea that \nc is the more fundamental measure for the gas-to-dust ratio than \nh.  

\end{enumerate}

Our study provides a new perspective regarding the physical underpinnings of the measurements of dust in our galaxy and beyond, and can be used to inform efforts to model extinction curves, grain size evolution, and attenuation in galaxies.

\acknowledgments

The authors would like to acknowledge the usefulness of the SciPy \citep{scipy} and \texttt{extinction} \citep{extinction.py} packages in Python.


\end{document}